\newcommand{\extended}[1]{}      

\documentclass[submission,copyright,creativecommons]{eptcs}
\providecommand{\event}{SR 2014} 
\usepackage{breakurl}             

\usepackage{amssymb}
\usepackage{amsmath}
\usepackage{amsthm}
\usepackage{comment}
\usepackage{color}
\usepackage[ruled]{algorithm}
\usepackage{algorithmic}
\usepackage{msc}
\usepackage{egameps}
\usepackage{xspace}
\usepackage{subfigure}

\definecolor{grey}{rgb}{0.6,0.6,0.6}

\newcommand{\remark}[2]{}

\newcommand{\defend}[1]{[\mathit{{#1}}]}

\newcommand{\powerset}{\mathcal{P}}

\newcommand{\sign}{\textsf{sign}}

\newcommand{\nash}{\textup{NE}\xspace}
\newcommand{\optnash}{\textup{OptNE}\xspace}

\newtheorem{theorem}{Theorem}[section]
\newtheorem{example}[theorem]{Example}
\newtheorem{proposition}[theorem]{Proposition}
\newtheorem{corollary}[theorem]{Corollary}
\newtheorem{definition}[theorem]{Definition}

\newenvironment{itemize2}{\begin{itemize}\itemsep 0in}{\end{itemize}}
\newenvironment{enumerate2}{\begin{enumerate}\itemsep 0in}{\end{enumerate}}

\newcommand{\set}[1]{\left\{#1\right\}}

\title{On Defendability of Security Properties}
\author{Wojciech Jamroga
\institute{\!\!\!\!\!\!Computer Science and Communication\\\& Interdisciplinary Centre for Security, Reliability, and Trust,\\University of Luxembourg}
\email{wojtek.jamroga@uni.lu}
\and
Matthijs Melissen
\institute{\!\!\!\!\!\!Computer Science and Communication,\\University of Luxembourg\\\& School of Computer Science,\\University of Birmingham}
\email{m.melissen@cs.bham.ac.uk}
\and
Henning Schnoor
\institute{Arbeitsgruppe Theoretische Informatik,\\University of Kiel}
\email{henning.schnoor@email.uni-kiel.de}
}
\def\titlerunning{On Defendability of Security Properties}
\def\authorrunning{W. Jamroga, M. Melissen, and H. Schnoor}

\begin{document}
\maketitle

\begin{abstract}
We study the security of interaction protocols when incentives of participants are taken into account.
We begin by formally defining correctness of a protocol, given a notion of rationality and utilities of participating agents.
Based on that, we propose how to assess security when the precise incentives are unknown.
Then, the security level can be defined in terms of \emph{defender sets}, i.e.,
sets of participants who can effectively ``defend'' the security property as long as they are in favor of the property.

We present some theoretical characterizations of defendable protocols under Nash equilibrium, first for bijective games (a standard assumption in game theory), and then for games with non-injective outcomes that better correspond to interaction protocols.
Finally, we apply our concepts to analyze fairness in the ASW contract-signing protocol.
\end{abstract}

\section{Introduction}

Interaction protocols are ubiquitous in multi-agent systems. Protocols can be modeled as games, since every participant in the protocol has several strategies that she can employ. From a game-theoretic perspective, protocols are an interesting class of games since they have a \emph{goal}, i.e., a set of outcomes that are preferred by the designer of the protocol. \emph{Security protocols} use cryptography to enforce their goals against any possible behavior of participants. Such a protocol is deemed correct with respect to its goal if the goal is achieved in all runs where a predefined subset of players follows the protocol.

We point out that this definition of correctness can be too strong, since violation of the goal may be achievable only by irrational responses from the other players.
On the other hand, the definition may also prove too weak when the goal can be only achieved by an irrational strategy of agents supporting the goal, in other words: one that they should never choose to play.
To describe and predict rational behavior of agents,
game theory has proposed a number of \emph{solution concepts}~\cite{Osborne94gamet}. Each solution concept captures some notion of rationality which may be more or less applicable in different contexts.
We do not fix a particular solution concept, but consider it to be a parameter of the problem.

Our main contributions are the following. First, in Section~\ref{sect:incentive correctness}, we define a parametrized notion of \emph{rational correctness} for security protocols, where the parameter is a suitable solution concept. Secondly, based on this notion, we define a concept of \emph{defendability of security} in a protocol, where the security property is guaranteed under relatively weak assumptions (Section~\ref{sec:defendability}).
Thirdly, in Section~\ref{sec:charcterizations}, we propose a \emph{characterization} of defendable security properties when rationality of participants is based on Nash equilibrium.  Finally, we consider the case of mixed strategies in Section~\ref{sect:mixed strategies}, we generalize the results to non-injective game models in Section~\ref{sec:non-injective}, and apply our concepts to analyze fairness in the ASW contract-signing protocol in Section~\ref{section:asw protocol}.
Most of this paper (Sections~\ref{sec:protocols}--\ref{sect:mixed strategies}) is a compressed version of the material already published in~\cite{Jamroga13defendable}. The novel contribution is presented in Sections~\ref{sec:non-injective} and~\ref{section:asw protocol}.

We want to emphasize that our work does not focus on ``classical'' security protocols where most participants are assumed to be ``honest'', i.e., to follow a typically deterministic sequence of actions.
More appropriately, we should say that we study \emph{interaction protocols} in general, where actions of participants may or may not be ``honest'', and the actual set of available behaviors depends on the execution semantics of the protocol.
We believe that the two kinds of assumptions (honesty vs.~being in favor of the protocol objective) are largely orthogonal.
A study of interplay between the two is left for future work.

\subsection{Related Work}

Researchers have considered protocol execution as a game with the very pessimistic assumption that the only goal of the other participants (``adversaries'') is to break the intended security property of the protocol.
In this case,
a protocol is correct if the ``honest'' participants have a strategy such that, for all strategies of the other agents, the goal of the protocol is satisfied (cf.~e.g.~\cite{Kremer02game}). Recently, protocols have been analyzed with respect to some game theoretic notions of rationality~\cite{Fuchsbauer10rational,Asharov11secure} where preferences of participants are taken into account.
An overview of connections between cryptography and game theory is given in~\cite{Dodis07cryptoGT}. Another survey~\cite{Moore11economics} presents arguments suggesting that study of incentives in security applications is crucial.
%
Butty\'an, Hubaux and \v{C}apkun \cite{Buttyan04formal} model protocols in a way similar to ours,
and also use incentives to model the behavior of agents.
However, they restrict their analysis to strongly Pareto-optimal Nash equilibria which is not necessarily a good solution concept for security protocols:
First, it is unclear why agents would \emph{individually} converge to a strongly Pareto-optimal play.
Moreover, in many protocols it is unclear why agents would play a Nash equilibrium in the first place.
Our method is more general, as we use the solution concept as a parameter to our analysis.
Asharov et al. (2011) \cite{Asharov11secure} use game theory to study gradual-release fair exchange protocols.
They consider a protocol to be game-theoretically fair if the strategy that never aborts the protocol is a computational Nash-equilibrium.
They prove that their analysis allows for solutions that are not admitted by the traditional cryptographic definition.
Groce and Katz \cite{Groce12fair} show that if agents have a strict incentive to achieve fair exchange,
then gradual-release fair exchange without trusted third party (TTP) is possible under the assumption that the other agents play rationally.
Syverson \cite{Syverson98weakly} presents a \emph{rational exchange} protocol for which he shows that ``enlightened, self-interested parties'' have no reason to cheat.
Finally, Chatterjee \& Raman \cite{Chatterjee10signing} use assume-guarantee synthesis for synthesis of contract signing protocols.

In summary, rationality-based correctness of protocols has been studied in a number of papers, but usually with a particular notion of rationality in mind. In contrast, we define a concept of correctness where a game-theoretic solution concept is a parameter of the problem. Even more importantly, our concept of \emph{defendability} of a security property is completely novel. The same applies to our characterizations of defendable properties under Nash equilibrium.

\section{Protocols and Games}\label{sec:protocols}

A protocol is a specification of how agents should interact.
Protocols can contain \emph{choice points} where several actions are available to the agents.
An agent is \emph{honest} if he follows the protocol specification,
and \emph{dishonest} otherwise, i.e., when he behaves in a way that is not allowed by the protocol.
In the latter case, the agent is only restricted by the physical and logical actions that are available in the environment.
For instance, in a cryptographic protocol, dishonest agents can do anything that satisfies properties of the cryptographic primitives, assuming perfect cryptography (as in \cite{Kremer03fairexchange}).
The protocol, together with a model of the environment of action, a subset of agents who are assumed to be honest, and the operational semantics of action execution, defines a multi-agent transition system that we call the \emph{model} of the protocol.
In the rest of the paper, we focus on protocol models, and abstract away from how they arise.
We also do not treat the usual ``network adversary'' that can intercept, delay and forge messages, but essentially assume the existence of secure channels. The issue of the ``network adversary'' is of course highly relevant for security protocols, but orthogonal to the aspects we discuss in this paper. In the full version of this paper~\cite{Jamroga13defendable}, we present contract signing protocols as a running example. In such a protocol, Alice and Bob want to sign a contract. Among the most relevant game-theoretic security properties of such protocols are fairness, balancedness, and abuse-freeness.

We use \emph{normal-form games} as abstract models of interaction in a protocol.

\begin{definition}[Frames and games]
A \emph{game frame} is a tuple $\Gamma = (N, \Sigma)$, where
$N = \{A_1, \dots, A_{|N|}\}$ is a finite set of \emph{agents},
 and $\Sigma = \Sigma_{A_1} \times \dots \times \Sigma_{A_{|N|}}$ is a set of strategy profiles\extended{ ($\Sigma_{A_i}$ is a strategy set for each $A_i \in N$)}.
\extended{
  We write $s_{-i}$ for the tuple of strategies by agents $N \backslash \{i\}$,
  and if $N = \{A_1, \ldots, A_n\}$, we write $(s'_{A_i},s_{A_{-i}})$ for the strategy profile
   $(s_{A_1}, \ldots, s_{A_{i-1}}, s'_{A_i}, s_{A_{i+1}}, \ldots s_{A_n})$.
}

A \emph{normal-form (NF) game} is a game frame plus a \emph{utility profile} $u = \{u_1, \dots, u_{|N|}\}$
where $u_i : \Sigma \rightarrow \mathbb{R}$ is a {utility function} assigning utility values to strategy profiles.
\end{definition}

Game theory uses \emph{solution concepts} to define which strategy profiles capture rational interactions.
Let $\mathcal{G}$ be a class of games with the same strategy profiles $\Sigma$.
Formally, a solution concept for $\mathcal{G}$ is a function $SC : \mathcal{G} \rightarrow \powerset(\Sigma)$ that, given a game, returns a set of \emph{rational} strategy profiles.
Well-known solution concepts include e.g. Nash equilibrium (\nash), dominant and undominated strategies, Stackelberg equilibrium, Pareto optimality etc.


\noindent\textbf{Protocols as Games.}
Let $P$ be a model of a protocol. We will investigate properties of $P$ through the game frame $\Gamma(P)$ in which strategies are \emph{conditional plans} in $P$, i.e., functions that specify for each choice point which action to take.
A set of strategies, one for each agent, uniquely determines a \emph{run} of the protocol, i.e., a sequence of actions that the agents will take.
$\Gamma(P)$ takes runs to be the outcomes in the game, and hence maps strategy profiles to runs.

Security protocols are designed to achieve one or more \emph{security requirements} and/or \emph{functionality requirements}.
We only consider requirements that can be expressed in terms of
single runs having a certain property.
We model this by a subset of possible behaviors, called the \emph{objective of the protocol}.%

\begin{definition}
Given a game frame $\Gamma = (N, \Sigma)$, an \emph{objective} is a set $\gamma \subseteq \Sigma$.
We call $\gamma$ \emph{nontrivial} in $\Gamma$ iff $\gamma$ is neither impossible nor guaranteed in $\Gamma$, i.e., $\emptyset\neq\gamma\neq\Sigma$.
\end{definition}

\section{Incentive-Based Security Analysis}\label{sec:analysis}

In this section, we give a definition of correctness of security protocols that takes into account rational decisions of agents,
based on their incentives.
\extended{First, we give a definition of correctness of protocols, assuming that the utility profile is known.
Then we proceed to the more general situation, where correctness of a protocol is to be verified without knowing the utility profiles of the agents.}
%

\subsection{Incentive-Based Correctness}\label{sect:incentive correctness}

As we have pointed out, the requirement that all strategy profiles satisfy the objective might be too strong.
Instead, we will require that all \emph{rational} runs satisfy the objective.
In case there are no rational runs, all behaviors are equally rational; then, we require that all strategy profiles must satisfy $\gamma$.
\begin{definition}\label{def:correctness}
A protocol model represented as game frame $\Gamma = (N, \Sigma)$ with utility profile $u$
is \emph{correct with respect to objective $\gamma$ under solution concept $SC$},
written $(\Gamma, u) \models_{SC} \gamma$,
iff:
\[ \left\{ \begin{array}{ll}
    SC(\Gamma,u)\subseteq\gamma\quad & \text{if }SC(\Gamma,u)\neq\emptyset \\
    \gamma=\Sigma       & \text{otherwise}.
\end{array}\right. \]
%
\end{definition}

\subsection{Unknown Incentives}\label{sect:correctness definition}

Definition~\ref{def:correctness} applies to a protocol when a utility profile is given.
However, the exact utility profiles are often unknown.
One way out is to require the protocol to be correct for \emph{all possible} utility profiles.

\begin{definition}
A protocol model represented by game frame $\Gamma$ is \emph{valid} with respect to objective $\gamma$ under solution concept $SC$
(written $\Gamma \models_{SC} \gamma$) iff $(\Gamma, u) \models_{SC} \gamma$ for all utility profiles $u$.
%
\end{definition}

It turns out that, under some reasonable assumptions, protocols are only valid for trivial objectives.
\begin{definition}
Let $G = (N, \Sigma, (u_1,\ldots, u_n))$.
Let $\pi = (\pi_1, \ldots, \pi_n)$, where for all $i \in N$, $\pi_i : \Sigma_i \rightarrow \Sigma_i$ is a permutation on $\Sigma_i$.
We slightly abuse the notation by writing $\pi((s_1, \ldots, s_n))$ for $(\pi_1(s_1), \ldots, \pi_n(s_n))$.
A solution concept is \emph{closed under permutation} iff
$s \in SC((N, \Sigma, (u'_1,\ldots, u'_n)))$ if and only if $\pi(s) \in SC((N, \Sigma, (u'_1\circ\pi_1^{-1},\ldots,u'_n\circ\pi_n^{-1})))$.
\end{definition}
%

\begin{theorem}
\label{the:permutation}
If $SC$ is closed under permutation, then $\Gamma \models_{SC} \gamma$ iff $\gamma = \Sigma$.\footnote{
  For proofs of all theorems and definitions of auxiliary concepts, we refer to the original paper~\cite{Jamroga13defendable}. }
\end{theorem}
%

Thus, correctness for all distributions of incentives is equivalent to correctness in all possible runs.
%

\subsection{Defendability of Protocols}\label{sec:defendability}

Typical analysis of a protocol implicitly assumes some participants to be aligned with its purpose.
E.g., one usually assumes that communicating parties are interested in exchanging a secret without the eavesdropper getting hold of it, that a bank wants to prevent web banking fraud etc.
In this section, we formalize this idea by assuming a subset of agents, called the \emph{defenders} of the protocol, to be in favor of its objective.
Our new definition of correctness says that a protocol is correct with respect to some objective $\gamma$ if and only if it is correct with respect to every utility profile in which the preferences of all defenders comply with $\gamma$.\footnote{
  There is an analogy of the concept to~\cite{Agotnes10robust} where ``robust'' goals are studied, i.e., goals that are achieved as long as a selected subset of agents behaves correctly. }

\begin{definition}
A group of agents $D \subseteq N$ \emph{supports} the objective $\gamma$ in game $(N, \Sigma, u)$ iff for all $i \in D$, if $s \in \gamma$ and $s' \in \Sigma \setminus \gamma$ then $u_i(s) > u_i(s')$.

A protocol model represented as game frame $\Gamma$ is \emph{defended by agents $D$},
written $\Gamma \models_{SC} \defend{D} \gamma$,
iff $(\Gamma, u) \models_{SC} \gamma$ for all utility profiles $u$ such that $D$ supports $\gamma$ in game $(\Gamma,u)$.
%
\end{definition}
%
Clearly, if there are no defenders, then defendability is equivalent to ordinary protocol validity:
\begin{proposition}
\label{the:nodefenders}
If $\Gamma$ is a game frame and $SC$ is a solution concept, we have that $\Gamma \models_{SC} \defend{\emptyset} \gamma$ iff $\Gamma \models_{SC} \gamma$.
\end{proposition}
%

If all agents are defenders, any protocol is correct, as long as the solution concept does not select \emph{strongly Pareto-dominated} strategy profiles, and there always is some strategy profile which is rational according to the solution concept.
\extended{
  \remark{HS}{This does not include Nash equilibria, correct? So do we know an interesting example where the theorem
  fails for \nash? (This is probably too late for AAMAS, but we could keep this in mind.}

  We formalize this in Th.~\ref{the:pareto}.

  \begin{definition}
  A strategy profile $s \in \Sigma$ is Pareto-dominated
  if there exists $s' \in \Sigma$ such that $u_i(s') > u_i(s)$ for all $i \in N$.
  \end{definition}
}
\begin{definition}
A solution concept is \emph{weakly Pareto} iff it never selects a strongly Pareto dominated outcome (i.e., such that there exists another outcome strictly preferred by all the players).
It is \emph{efficient} iff it never returns the empty set.
\end{definition}

\begin{theorem}\label{the:pareto}
If $\Gamma$ is a game frame and $SC$ is an efficient weakly Pareto solution concept then $\Gamma \models_{SC} \defend{N} \gamma$.
\end{theorem}


Many solution concepts are both efficient and weakly Pareto, for example: Stackelberg equilibrium, maximum-perfect cooperative equilibrium, backward induction and subgame-perfect Nash equilibrium in perfect information games.
%
On the other hand, Nash equilibrium is neither weakly Pareto nor efficient\extended{ (the Hi/Lo from game Figure~\ref{sfig:hilo} being a counterexample)},
and equilibrium in dominant strategies is weakly Pareto but not necessarily efficient.

Clearly, defendability of a protocol is monotonic with respect to the set of defenders.
This justifies the following definition.
\begin{definition}
The \emph{game-theoretic security level} of protocol $P$ is the antichain of minimal sets of defenders that make the protocol correct.
\end{definition}
%

\section{Characterizing Defendability under Nash Equilibrium}\label{sec:charcterizations}

In this section, we turn to properties that can be defended if agents' rationality is based on Nash equilibrium or Optimal Nash Equilibrium.

\subsection{Defendability under Nash Equilibrium}\label{sec:defendability-ne}


From Theorem~\ref{the:permutation}, we know that no protocol is valid under Nash equilibrium (\nash) for any nontrivial objective, since \nash\ is closed under permutation. Do things get better if we assume some agents to be in favor of the security objective? We now look at the extreme variant of the question, i.e., defendability by the grand coalition $N$. Note that, by monotonicity of defendability wrt the set of defenders $D$, nondefendability by $N$ implies that the objective is not defendable by any coalition at all.

Our first result in this respect is negative: we show that in every game frame there are nontrivial objectives that are not defendable under \nash.
\begin{theorem}
Let $\Gamma$ be a game frame with at least two players and at least two strategies per player. Moreover, let $\gamma$ be a singleton objective, i.e., $\gamma=\{\omega\}$ for some $\omega\in\Sigma$. Then, $\Gamma\not\models_{\nash}\defend{N}\gamma$.
\end{theorem}
%

%
%

In particular, the construction from the above proof shows that, as mentioned before, there are cases where the ``defending'' coalition has a strategy to achieve a goal $\gamma$, but there are still rational plays in which the goal is not achieved.

To present the general result that characterizes defendability of security objectives under Nash equilibrium, we need to introduce additional concepts. In what follows, we use $s[t_i/i]$ to denote $(s_1,\dots,s_{i-1},t_i,\linebreak s_{i+1},\dots,s_N)$, i.e., the strategy profile that is obtained from $s$ when player $i$ changes her strategy to $t_i$.

\begin{definition}
Let $\gamma$ be a set of strategy profiles in $\Gamma$. The \emph{deviation closure} of $\gamma$ is defined as $Cl(\gamma) = \{s\in\Sigma \mid \exists i\in N, t_i\in\Sigma_i\ .\ s[t_i/i]\in\gamma \}$.
\end{definition}

$Cl(\gamma)$ extends $\gamma$ with the strategy profiles that are reachable by unilateral deviations from $\gamma$. Thus, $Cl(\gamma)$ can be seen as the closure of $\gamma$ with the behaviors that are relevant for Nash equilibrium. Moreover, the following notion captures strategy profiles that can be used to construct sequences of unilateral deviations ending up in a cycle.

\begin{definition}
A \emph{strategic knot} in $\gamma$ is a subset of strategy profiles $S\subseteq\gamma$ such that there is a permutation $(s^1,\dots,s^k)$ of $S$ where:\ (a) for all $1\le j<k$, $s^{j+1}= s^j[s^{j+1}_i/i]$ for some $i\in N$, and\ (b) $s^j= s^k[s^{j}_i/i]$ for some $i\in N, j<k$.
\end{definition}

Essentially, this means that every strategy $s^{j+1}$ is obtained from $s^j$ by a unilateral deviation of a single agent. If these deviations are rational (i.e., increase the utility of the deviating agent), then the knot represents a possible endless loop of rational, unilateral deviations which precludes a group of agents from reaching a stable joint strategy. We now state the main result of this section.

\begin{theorem}\label{prop:defendNE}
Let $\Gamma$ be a finite game frame\extended{ with $N\ge 2, |\Sigma_1|\ge 3, |Sigma_2|\ge 3$,}
and $\gamma$ a nontrivial objective in $\Gamma$. Then, $\Gamma\models_{\nash}\defend{N}\gamma$\ iff\ $Cl(\gamma)=\Sigma$ and there is a strategy profile in $\gamma$ that belongs to no strategic knots in $\gamma$.
\end{theorem}

\extended{
  [Add some examples of defendable patterns + computational complexity of verifying if $\gamma$ is $N$-defendable in $\Gamma$ under $\nash$]
}

\subsection{Optimal Nash Equilibria}

Nash equilibrium is a natural solution concept for a game played repeatedly until the behavior of all players converges to a stable point.
For a one-shot game, \extended{the situation is more delicate. In that case, }\nash possibly captures convergence of the process of deliberation\extended{ (about what the others might want to play, what they think I plan to play, etc.)}. It can be argued that, among the available solutions, no player should contemplate those which are strictly worse for everybody when compared to another stable point.
This gives rise to the following refinement of Nash equilibrium: $\optnash(\Gamma,u)$ is the set of \emph{optimal Nash equilibria} in game $(\Gamma,u)$, defined as those equilibria \emph{that are not strongly Pareto-dominated by another Nash equilibrium}.
Defendability by the grand coalition under $\optnash$ has the following simple characterization.

\begin{theorem}\label{prop:defendOptNE}
Let $\Gamma$ be a finite game frame\extended{ with $N\ge 2, |\Sigma_1|\ge 3, |Sigma_2|\ge 3$,}
and $\gamma$ a nontrivial objective in $\Gamma$. Then, $\Gamma\models_{\optnash}\defend{N}\gamma$ iff there is a strategy profile in $\gamma$ that belongs to no strategic knots in $\gamma$.
\end{theorem}
%
%


\section{Defendability in Mixed Strategies}\label{sect:mixed strategies}

So far, we considered only deterministic (pure) strategies. It is well known that for many games and solution concepts, rational strategies exist only when taking mixed strategies into account. We now extend our definition of correctness to mixed strategies, i.e., randomized conditional plans represented by probability distributions over pure strategies from $\Sigma_{A_i}$.
Let $dom(s)$ be the support (domain) of a mixed strategy profile $s$, i.e., the set of pure strategy profiles that have nonzero probability in $s$. We extend the notion to sets of mixed strategy profiles in the obvious way. By $SC^m$ we denote the variant of $SC$ in mixed strategy profiles.
A protocol is correct in mixed strategies iff all the possible behaviors resulting from a rational (mixed) strategy profile satisfy the goal $\gamma$; formally: $\Gamma,u\models_{SC}^m\gamma$ iff $dom(SC^m(\Gamma,u)) \subseteq \gamma$ when $SC^m(\Gamma,u) \neq \emptyset$ and $\gamma=\Sigma_\Gamma$ otherwise.
The definitions of protocol validity and defendability in mixed strategies ($\Gamma\models_{SC}^m\gamma$ and $\Gamma\models_{SC}^m\defend{D}\gamma$) are analogous.
For defendability in mixed strategies under Nash equilibrium, we have the following, rather pessimistic result.
\begin{theorem}\label{prop:defendNE-mixed}
Let $\Gamma$ be a finite game frame, and $\gamma$ an objective in it. Then, $\Gamma,u\models_{\nash}^m\defend{N}\gamma$\ iff\ $\gamma=\Sigma$.
\end{theorem}
%
%
%
%

On the other hand, it turns out that \emph{optimal Nash equilibrium} yields a simple and appealing characteristics of $N$-defendable properties. In the following, $\gamma$ is closed under convex combination of strategies iff every combination of strategies that appear in some profile in $\gamma$ again is an element of $\gamma$.
\begin{theorem}\label{prop:defendOptNE-mixed}
$\Gamma\models_{\optnash}^m\defend{N}\gamma$\ iff\ $\gamma=Conv(\gamma)$, i.e., $\gamma$ is closed under convex combination of strategies.
\end{theorem}
%
%

\begin{corollary}
$\Gamma\models_{\optnash}^m\defend{N}\gamma$\ iff\ there exist subsets of individual strategies \linebreak $\chi_1\subseteq\Sigma_1, \dots, \chi_{|N|}\subseteq\Sigma_{|N|}$ such that  $\gamma=\chi_1\times\dots\times\chi_{|N|}$.
\end{corollary}
That is, security property $\gamma$ is defendable by the grand coalition in $\Gamma$ iff $\gamma$ can be \emph{decomposed into constraints on individual behavior of particular agents}.

\begin{figure}[t]
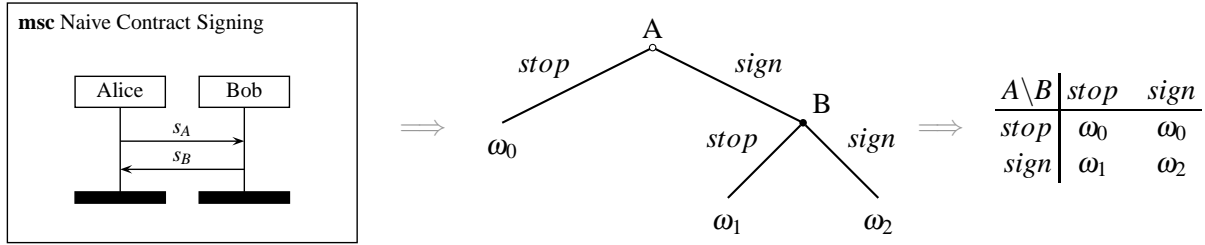

  \centerline{
    \begin{tabular}{@{}c@{}c@{}c@{}c@{}c@{}}
    \begin{tabular}{c}
      \psscalebox{.75}{
      \input{naive-signing.tex}
      }
    \end{tabular}
     &
    \begin{tabular}{c@{\!\!\!\!}}
      {\color{grey} $\Longrightarrow$}
    \end{tabular}
     &
    \begin{tabular}{@{}c}
      \input{naive-EFgame.tex}
    \end{tabular}
     &
    \begin{tabular}{c}
      {\color{grey} $\Longrightarrow$}
    \end{tabular}
     &
    \begin{tabular}{c}
      \begin{tabular}{@{\,}r@{\ }|@{\ }c@{\quad}c@{\,}}
  $A \backslash B$ & $stop$ & $sign$ \\ \hline
  $stop$ & $\omega_0$ &  $\omega_0$\\
  $sign$ & $\omega_1$ & $\omega_2$
\end{tabular}

    \end{tabular}
    \end{tabular}
  }
\caption{Naive contract-signing: from protocol to EF game to NF game}
\label{fig:naive-games}
\end{figure}

\section{Defendability in Non-Injective Games}\label{sec:non-injective}

Normal game frames are usually defined in the literature as $\Gamma = (N,\Sigma,\Omega,o)$, where $N,\Sigma$ are as before, $\Omega$ is the set of (abstract) \emph{outcomes} of the game, and $o : \Sigma \rightarrow \Omega$ maps strategy profiles to outcomes.
Our analysis so far has been based on the standard assumption that $o$ is a bijection. In other words, there is a 1-1 relationship between joint {behaviors} of agents and the outcomes of those behaviors. Then, we can identify outcomes with strategy profiles, and omit the former from the game model.
However, the standard construction of a game model from a protocol assumes the outcomes to be \emph{runs} of the protocol. In that case, the assumption does \emph{not} hold; in particular, the mapping is not injective.
\begin{example}
Consider the naive contract signing protocol in Figure~\ref{fig:naive-games}. Alice sends her signature to Bob, who responds with his signature.
Alice and Bob can stop the protocol at any moment (thereby deviating from the protocol). If we assume {runs} of the protocol to be the outcomes, this gives rise to an Extensive Form game frame, which can be then transformed to an NF game frame by the canonical construction.
Clearly, the mapping between strategy profiles and outcomes is not injective.
%
\end{example}

In general NF games, utility functions assign utility values to \emph{outcomes} rather than strategy profiles. That is, $u_i : \Omega \rightarrow \mathbb{R}$. Moreover, an objective is assumed to select a \emph{subset of outcomes}. This follows from the methodological assumption that an outcome encapsulates every relevant aspect of the play that has occurred.
We observe that the definitions in Section~\ref{sec:analysis} can be lifted to the general case by changing the types of $u_i$ and $\gamma$ accordingly.
However, the results in Sections~\ref{sec:charcterizations}--\ref{sect:mixed strategies} cannot be lifted that easily.
Games with non-injective outcome functions require a more general treatment, which we present below.

\begin{definition}
Given a game frame $\Gamma$, we define the \emph{deviation graph of $\Gamma$ ($Dev(\Gamma)$)} to be the undirected graph where outcomes from $\Gamma$ are vertices, and edges connect outcomes
that are obtained from strategy profiles which differ only in $1$ individual strategy (thus corresponding to a potential unilateral deviation).

Moreover, for an objective $\gamma\subseteq\Omega$, we will use $Dev_\gamma(\Gamma)$ to denote the subgraph of $Dev(\Gamma)$ consisting only of the vertices from $\gamma$ and the edges between them.
\end{definition}

It is easy to see that the construction of $Dev(\Gamma)$ and $Dev_\gamma(\Gamma)$ from $\Gamma,\gamma$ is straightforward.
Let $V$ be a subset of nodes in a graph. We define the \emph{neighborhood of $V$}, denoted $Neighb(V)$, as $V$ together with all the nodes adjacent to $V$.
We observe that $Neighb(V)$ ``implements'' the deviation closure of $V$ in $Dev(\Gamma)$. Moreover, $\omega$ does not lie on a strategic knot iff its connected component does not include a cycle.
This leads to the following, more general, characterizations of defendability (we omit the proofs due to lack of space).
Again, we assume that $\gamma$ is nontrivial, i.e., $\emptyset\neq\gamma\neq\Omega$.

\begin{theorem}
$\gamma$ is defended by the grand coalition in $\Gamma$ under Nash equilibrium iff:
\begin{enumerate2}
\item The neighborhood of $\gamma$ in in $Dev(\Gamma)$ covers the whole graph ($Neighb(\gamma) = \Omega$), and
\item $Dev_\gamma(\Gamma)$ includes at least one connected component with no cycles.
\end{enumerate2}
\end{theorem}

\begin{theorem}
$\gamma$ is defended by the grand coalition in $\Gamma$ under optimal Nash equilibrium iff
$Dev_\gamma(\Gamma)$ includes at least one connected component with no cycles.
\end{theorem}

\begin{theorem}\label{prop:defendOptNE-mixed-noninjective}
$\gamma$ is defended in mixed strategies by the grand coalition in $\Gamma$ under optimal Nash equilibrium iff
$\gamma$ is obtained by a convex combination of strategies.
\end{theorem}

\newcommand{\ASW}{\ensuremath{P_{\mathrm{ASW}}}\xspace}

\section{Example: The ASW contract-signing protocol}\label{section:asw protocol}

A contract-signing protocol is used by two participants, usually called Alice and Bob, to sign a contract over an asymmetric medium as the internet. The central security properties are \emph{fairness} (Alice should get a signed copy of the contract if and only if Bob gets one), \emph{balancedness} (there is no point in the protocol run where Bob alone can decide whether the contract will be signed or not, i.e., Alice cannot abort the signing anymore but Bob still can abort) and \emph{abuse-freeness} (if balance cannot be achieved, then at least Bob should not be able to prove the fact that he has the above-mentioned strong position in the current state of the protocol to an outsider). The contract-signing protocol \ASW, introduced in~\cite{Asokan98asynchronous}, uses \emph{commitments}, which are legally binding ``declarations of intent'' by Alice and Bob to sign the contract.
The protocol operates as follows:\
(1) Alice sends a {commitment} $cm_A$ to Bob;\
(2) Bob sends his commitment $cm_B$ to Alice;\
(3) Alice sends the contract $sc_A$, digitally signed with her signature, to Bob;\
(4) Bob sends the contract $sc_B$, signed with his signature, to Alice.

If one of these messages is not sent by the corresponding signer, the other party may contact the TTP:
\begin{itemize2}
 \item If Alice does not receive a commitment from Bob, she can contact the TTP with an \emph{abort request}, which instructs the TTP to mark this session of the protocol as aborted;
 \item If Bob does not receive Alice's signature, but has her commitment, he can send a \emph{resolve request} to the TTP, who then issues a \emph{replacement contract} (a document that is legally equivalent to the contract signed by Alice), unless Alice has sent an abort request earlier,
 \item If Alice does not receive Bob's signature, but has his commitment, she can send a \emph{resolve request} to the TTP as well, which allows her to receive a replacement contract.
\end{itemize2}

It can be shown that the protocol is fair if the TTP is reliable (it will never stop the protocol on its own). It is also {balanced} if neither Alice nor Bob can drop or delay messages from the other signer to the TTP.
Let us denote outcomes by sets of agents who have obtained the signature of the other player. Thus, $\emptyset$ represents the situation where nobody got a signed contract, $\{\sign_A\}$ the situation where Alice obtained Bob's signature but note vice versa, etc.
Applying the definitions in Section~\ref{sec:defendability}, one can show the following. If $SC$ is either \emph{Nash equilibrium} or \emph{undominated strategies}, we have:
\begin{enumerate2}
\item $\ASW\models_{SC}\defend{\{\mathrm{Bob}\}}\{\emptyset,\{\sign_B\},\{\sign_A,\sign_B\}\}$,
\item $\ASW\models_{SC}\defend{\{\mathrm{Alice}\}}\{\emptyset,\{\sign_A\},\{\sign_A,\sign_B\}\}$.
\end{enumerate2}

We now consider the case where TTP is not necessarily reliable. If the TTP can stop the protocol at any time, then the protocol does not guarantee fairness anymore. On the other hand, if Bob wants the protocol to be fair, then he can ensure fairness by simply sending a signed contract to Alice as soon as he receives her signature. Clearly, Alice alone (without an honest TTP to assist her) cannot achieve fairness. Hence the game-theoretic security level of the ASW protocol without reliable TTP is the set $\set{\set{\mathrm{Bob}},\set{\mathrm{TTP}}}$. This holds for both Nash equilibrium and undominated strategies.

\section{Conclusions}

We propose a framework for analyzing security protocols (and other interaction protocols), that takes into account the incentives of agents.
In particular, we consider a novel notion of \emph{defendability} that guarantees that all the runs of the protocol are correct as long as a given subset of the participants (the ``defenders'') is in favor of the security property.
We have obtained some characterization results for defendability under Nash equilibria and optimal Nash equilibria. In the original paper~\cite{Jamroga13defendable}, we also address the computational complexity of the corresponding decision problems, both in the generic case and in some special cases.
In the future, we plan to combine our framework with results for protocol verification using game logics (such as ATL), especially for those solution concepts that can be expressed in that kind of logics.

\smallskip\noindent\textbf{Acknowledgements.} We thank the SR2014 reviewers for their extremely useful remarks. Addressing the fundamental ones was not possible in this extended abstract due to space and time constraints, but we will use them in the journal version of the paper (in preparation).

Wojciech Jamroga acknowledges support of the National Research Fund Luxembourg (FNR) under project GaLOT -- INTER/DFG/12/06.

\bibliographystyle{eptcs}
\bibliography{wojtek,wojtek-own}

\end{document}